\documentclass[12pt]{article}
\usepackage{amssymb,textcomp,amsmath,graphicx}
\begin{document}
\renewcommand{\thefootnote}{\fnsymbol{footnote}}
\newcommand{\be}{\begin{equation}}
\newcommand{\ee}{\end{equation}}
\newcommand{\tanb}{\mbox{$\tan \! \beta$}}
\newcommand{\cotb}{\mbox{$\cot \! \beta$}}
\newcommand{\cosb}{\mbox{$\cos \! \beta$}}
\newcommand{\sinb}{\mbox{$\sin \! \beta$}}
\newcommand{\wt}{\widetilde}
\def\gsim{\:\raisebox{-0.5ex}{$\stackrel{\textstyle>}{\sim}$}\:}
\def\lsim{\:\raisebox{-0.5ex}{$\stackrel{\textstyle<}{\sim}$}\:}

\pagestyle{empty}

\begin{flushright}
October 2012\\
\end{flushright}
\vspace*{1.5cm}

\begin{center}
{\Large \bf A Supersymmetric Explanation of the Excess of Higgs--Like Events
  at the LHC and at LEP} \\
\vspace{1cm}
{\large Manuel Drees} \\
\vspace*{6mm}
{\it Bethe Center for Theoretical Physics and Physikalisches
  Institut d. Universit\"at Bonn,\\ Nussallee 12, 53115 Bonn, 
  Germany} 
\end{center}

\begin{abstract}
The LHC collaborations have recently announced evidence for the
production of a ``Higgs--like'' boson with mass near 125 GeV. The
properties of the new particle are consistent (within still quite
large uncertainties) with those of the Higgs boson predicted in the
Standard Model (SM). This discovery comes nearly ten years after a
combined analysis of the four LEP experiments showed a mild excess of
Higgs--like events with a mass near 98 GeV. I show that both groups of
events can be explained simultaneously in the minimal supersymmetric
extension of the SM, in terms of the production and decay of the two
neutral CP--even Higgs bosons predicted by this model, and explore the
phenomenological consequences of this explanation. 

\end{abstract}
\clearpage

\setcounter{page}{1}

\pagestyle{plain}

\section{Introduction}

Recently the LHC collaborations ATLAS and CMS announced the discovery
of a ``Higgs--like'' boson with mass near 125 GeV
\cite{LHC_Higgs}. This new boson has been detected at the LHC chiefly in the
$\gamma \gamma$ and four lepton final states. In addition, there is
evidence, at the $\sim 3 \sigma$ level, for decays into $b \bar b$
pairs from the Tevatron experiments CDF and D0 \cite{TEV_Higgs}. 

Within the still quite large experimental (and theoretical
\cite{abro}) uncertainties, the properties of the new boson are
consistent with those of the single physical Higgs boson in the
Standard Model (SM). However, as well known, the scalar sector of the
SM is technically unnatural, since it suffers from quadratic
divergencies. These divergencies are canceled in supersymmetric
extensions of the SM \cite{hierarchy}. 

Even the simplest such theory, the Minimal Supersymmetric extension of
the SM (MSSM), contains two Higgs doublets, the second doublet being
required both for the cancellations of anomalies (from the higgsinos),
and in order to give masses to all quarks \cite{book}. As a result,
the MSSM contains three neutral physical Higgs states. In the absence
of CP violation, these can be classified as two CP--even states $h, H$
(with $m_h < m_H$) and one CP--odd state $A$. Most interpretations of
the new boson discovered by the LHC experiments (many of which were
published after the first experimental evidence was announced in
December 2011) within the MSSM focus on the possibility that it is the
lighter CP--even state $h$ \cite{MSSM_h,MSSM_both,mullah}. However,
achieving $m_h \simeq 125$ GeV is only possible if stop squarks are
very heavy. By most definitions, this requires a somewhat
uncomfortable amount of finetuning\footnote{See however
  ref.\cite{Barger} for an example spectrum with a 125 GeV Higgs boson
  and heavy stops nevertheless requiring little ``electroweak--scale
  finetuning''.} In this scenario the heavier neutral Higgs bosons $H,
A$ as well as the charged Higgs bosons $H^\pm$ can be essentially
arbitrarily heavy. In fact, in simple (constrained) scenarios of
supersymmetry breaking, the large lower bounds on (first generation)
squark masses typically require these states to be quite heavy; within
such constrained scenarios the new state therefore has to be
interpreted as $h$.

The possibility that instead the heavier state $H$ has been discovered
has also been entertained \cite{italians,MSSM_both,kajae}. In this
case the lighter CP--even state would obviously have to be lighter
than 125 GeV, and would need to satisfy limits from Higgs searches
both at the LHC \cite{LHC_Higgs} and at LEP \cite{leph}. As pointed
out in refs.\cite{MSSM_both, kajae} this is not difficult to achieve,
if $H$ is SM--like in agreement with experimental observations of the
new state at 125 GeV. In particular, sufficiently large branching
ratios for $H$ into four leptons, and into two photons, require the
couplings of $H$ to two massive gauge bosons to not differ very much
from the corresponding SM values. In the context of the MSSM this
automatically implies that $h$ has suppressed couplings to $W$ and
$Z$, making it difficult to detect.

Here I wish to point out that in this scenario $h$ might be put to
good use, by explaining an excess of Higgs--like events observed some
ten years ago by the four LEP collaborations
\cite{leph}.\footnote{Very recently it has been pointed out that both
  the LEP excess and the LHC discovery can be explained in the NMSSM
  \cite{begjks}, where the spectrum contains three CP--even and two
  CP--odd neutral Higgs bosons.} Actually, the combined LEP data
showed two regions of reconstructed Higgs mass where some excess
occurred. One was right at the kinematic limit, near 115 GeV. This
excess was observed mostly by the ALEPH collaboration \cite{aleph}; in
the combined data, its statistical significance reached only 1.7
standard deviations. It is compatible with an SM--like Higgs with this
mass. However, this interpretation is at odds with the interpretation
of the new particle discovered at the LHC as an SM--like Higgs boson.

The combination of the data from all four LEP experiments also
revealed \cite{leph} a somewhat more significant excess near 98 GeV,
with significance of about 2.3 standard deviations. This excess is
{\em not} compatible with an SM Higgs at that mass; rather, it's
compatible with an about ten times smaller production cross section
than predicted by the SM for this mass range. In the context of the
MSSM this can easily be arranged by reducing the $hZZ$ (and hence also
the $hWW$) coupling \cite{D1}. This implies that the other MSSM Higgs
bosons have to be relatively light: if they were heavy, $h$ would
become SM--like. A detailed analysis found an upper bound on $m_H$ of
about 140 GeV \cite{D1}. The new particle at 125 GeV therefore falls
right in the middle of the allowed range for $m_H$ in this scenario,
where the lower bound (of about 114 GeV) comes from LEP Higgs
searches.\footnote{Ref.\cite{italians} also considered the MSSM with
  $m_H \simeq 125$ GeV, including scenarios with $m_h \leq 110$ GeV
  and suppressed $ZZh$ couplings, in the context of an analysis of
  scenarios with a light neutralino as Dark Matter candidate, based on
  very early, preliminary LHC results. No bounds on the strengths of
  the $H$ signals were imposed, and $h$ was not required to explain
  the LEP excess near 98 GeV.}

The LHC discovery obviously greatly constrains the allowed parameter
space of this scenario, where now the masses of {\em both} CP--odd
Higgs bosons are fixed within a few GeV theoretical and experimental
uncertainty. However, as well known the MSSM Higgs sector is subject
to large radiative corrections. This introduces several new
parameters, the most important ones being the ones appearing in the
stop mass matrix. Here I present a detailed analysis of this
scenario. This not only updates ref.\cite{D1} by including the
constraint $m_H \simeq 125$ GeV; I also carefully compute the relevant
decay widths, and resulting branching ratios and signal strengths, of
the neutral Higgs bosons, where the latter are normalized to the
signal strength of the SM Higgs boson. I also include constraints from
null searches for neutral MSSM Higgs bosons decaying into tau pairs
performed by CMS \cite{cms_tau}, and for charged Higgs bosons produced
in top quark decays performed by ATLAS \cite{atlas_top}. The former
are more important, considerably limiting the allowed parameter space
of this scenario. I nevertheless find that $m_A$ masses roughly
between 95 and 150 GeV are allowed in this scenario. The charged Higgs
boson mass can reach up to about 170 GeV. The signals for $H$
production in both the di--photon and four lepton channels can be
considerably enhanced, but the {\em ratio} of these two signals cannot
exceed its SM prediction by more than about 35\%.

This analysis has been performed in the framework of the general MSSM,
where all relevant parameters are fixed directly at the weak (or TeV)
scale. In order to limit the size of the parameter space, I will not
specify soft breaking parameters for the first two generations of
sfermions, which have almost no impact on the masses and couplings of
Higgs bosons. This approach also permits me to ignore all constraints
from flavor physics, which depend very strongly on the flavor
structure of the soft breaking terms.

The rest of this paper is organized as follows. In Sec.~2 I describe
details of the analysis. In Sec.~3 I explore the parameter space that
is compatible with this explanation; in particular, I give allowed
ranges for physical quantities of interest, and explore correlations
between them. Finally, Sec.~4 contains a brief summary and conclusions.

\section{Details of the Analysis}

This analysis is performed in the framework of the general MSSM, where
all relevant parameters are fixed at the weak scale, and no
high--scale constraints on the spectrum of superpartners are imposed.

Obviously at least the leading radiative corrections \cite{radcorr1}
to the masses and mixing angle of the MSSM Higgs bosons have to be
included in any quantitative analysis. This is most easily done using
the effective potential (or, equivalently, Feynman diagrammatic
calculations with vanishing external momentum). Recall that the entire
Higgs spectrum should be relatively light in this scenario; this should
increase the reliability of the effective potential method.

In order to allow an efficient sampling of the parameter space, I only
include corrections from the top--stop and bottom--sbottom sectors to
the Higgs boson mass matrices, which give the by far most important
contributions. The relevant expressions have been taken from
refs.\cite{dn} (for the pure Yukawa corrections to both the neutral
and charged Higgs boson mass matrices); \cite{cdl} (for the mixed
electroweak--Yukawa corrections to the neutral Higgs boson mass
matrix); and \cite{dn,hhh} and \cite{hhh,carena} (for the inclusion of
leading higher order QCD and top Yukawa corrections, respectively, by
using running quark masses defined at the appropriate scale in the
one--loop effective potential). The leading SUSY QCD corrections are
included through the gluino--stop and gluino--sbottom corrections to
the top and bottom mass, respectively \cite{youichi}; as shown in
ref.\cite{heinemeyer}, this reproduces the full SUSY QCD correction
very accurately. The calculation performed here should reproduce the
neutral MSSM Higgs masses with an error of about two or three GeV
\cite{herror,heinemeyer}. This theoretical uncertainty will be
included in the constraints imposed on $m_h$ and $m_H$. Note that the
SUSY QCD corrections to the bottom mass are also included in the
calculation of the corresponding Yukawa couplings, which affect both
the partial widths of the neutral Higgs bosons into $b \bar b$ pairs
and the $b$ loop contribution to the partial widths of the decays into
gluon and photon pairs.

The running top mass, $m_t(m_t)$ is fixed to $165$ GeV (in the
$\overline{\rm DR}$ scheme). This corresponds to a pole mass near 173
GeV, the current central value \cite{topmass}. I also fix $m_b(m_b) =
4.25$ GeV. As final simplification, I have taken the soft breaking
parameters in the stop and sbottom mass matrices to be the same.  This
is always true for the masses of the superpartners of the left--handed
squarks, due to $SU(2)$ invariance, but the masses of the $SU(2)$
singlet squarks as well as the two $A-$parameters could in principle
be different. However, we will see that the CMS di--tau search
requires the ratio of vacuum expectation values $\tan\beta$ to be
relatively small, below $13$; as a result, sbottom loops are always
subdominant, and thus need not be treated as carefully as stop loops.

The most convincing signals for the new state at 125 GeV have been
found in the di--photon channel. This is obviously only accessible
through loop diagrams. In addition to the diagrams involving $W$
bosons or third generation fermions, diagrams involving charged Higgs
bosons as well as all third generation sfermions are included. It had
been noticed \cite{stauloop,kajae} that loops involving stau sleptons could
significantly change the di--photon widths of neutral CP--even MSSM
Higgs bosons. I therefore allow the $\tilde \tau_{L,R}$ soft breaking
masses as well as the trilinear soft breaking parameter $A_\tau$ to
vary independently from the parameters of the stop sector. However, it
turns out that in the given scenario, stau loops always make very
small contributions. Similarly, the partial widths of the Higgs bosons
into gluons are computed including loops of third generation quarks as
well as squarks (squark loops are absent in case of the CP--odd Higgs
boson). The relevant expressions are taken from \cite{abdel}.

Altogether we are thus left with ten free parameters: $\tan\beta, \,
m_A, \, \mu, \, m_{\tilde t_L}, \, m_{\tilde t_R}, \, A_t, \,
m_{\tilde \tau_L}, \, m_{\tilde \tau_R}, \, A_\tau, \, m_{\tilde
  g}$. This ten--dimensional parameter space has been scanned
randomly, subject to the following constraints not involving Higgs bosons:
\begin{subequations} \label{cons}\begin{align} 
|\mu|, \, m_{\tilde t_R}, \, m_{\tilde t_L}, \, m_{\tilde g}, \,
m_{\tilde \tau_L}, \, m_{\tilde \tau_R} &  \leq 5 \ {\rm TeV}; \\ 
|\mu|, \, m_{\tilde t_1}, \, m_{\tilde b_1}, \, m_{\tilde \tau_1}
& \geq 100 \ {\rm GeV}; \\ 
|m_{\tilde t_1} - m_{\tilde b_1} | &\leq 50 \ {\rm GeV} \ {\rm or} \
\max(m_{\tilde t_1}, m_{\tilde b_1}) > 300 \ {\rm GeV}; \\
m_{\tilde g} & \geq 600 \ {\rm GeV}; \\
|A_t|, |\mu|  & \leq 1.5 \left( m_{\tilde{t}_R} + m_{\tilde t_L}
\right) ; \\
|A_\tau|, |\mu|  & \leq 1.5 \left( m_{\tilde \tau_R} + m_{\tilde \tau_L} \right); \\
\delta \rho_{\tilde t \tilde b} & \leq 2 \cdot 10^{-3}.
\end{align} \end{subequations}
The first of these constraints is a (quite conservative) naturalness
criterion. Conditions (\ref{cons}b) ensure that higgsino--like
charginos (with mass $\sim |\mu|$) as well as the lighter physical
stop $(\tilde t_1)$, sbottom $(\tilde b_1)$ and stau ($\tilde \tau_1$)
states escaped detection at LEP \cite{pdg}. Condition (\ref{cons}c)
ensures that only one of the two lighter squark states can be below
300 GeV, unless they are close in mass. In the latter case they could
both be close in mass to the lightest neutralino, in which case
$\tilde t_1$ and $\tilde b_1$ pair production would lead to events
with a small amount of visible energy, which are difficult to detect.
Condition (\ref{cons}d) is a rather conservative interpretation of
gluino search limits in the general MSSM. Note that loops involving
gluinos affect the Higgs masses and mixing angle only at two--loop
order, but modify the $h b \bar b$ and $H b \bar b$ couplings already
at one--loop. The upper bounds (\ref{cons}e,f) on the parameters
determining mixing in the stop and stau sectors have been imposed to
avoid situations where $\tilde t$ or $\tilde \tau$ fields have
non--vanishing VEVs in the absolute minimum of the scalar potential
\cite{CCB}.\footnote{Note that ref.\cite{stauloop}, where the
  influence of light staus on the two--photon widths of MSSM Higgs
  bosons was first explored, only imposes the much weaker constraint
  on $|\mu|$ that follows from the requirement that the zero
  temperature tunnel rate into the false vacuum \cite{hisano_tunnel}
  is sufficiently small.} Finally, (\ref{cons}g) requires the
contribution of stop--sbottom loops to the electroweak $\rho$
parameter \cite{delrho} to be sufficiently small.

In order to be able to describe the (mild) excess of Higgs--like
events at LEP, and the properties of the new boson discovered at the
LHC, the Higgs sector has to simultaneously satisfy the following
constraints:
\begin{subequations} \label{h_cons} \begin{align}
95 \ {\rm GeV} &\leq m_h \leq 101 \ {\rm GeV}\,; \\
123 \ {\rm GeV} &\leq m_H \leq 128 \ {\rm GeV}\,; \\
0.056 &\leq \sin^2(\alpha - \beta) \leq 0.144\,; \\
0.5 &\leq R_H^{VV} \leq 2.0\ \ \ (V=W,Z)\,; \\
0.5 &\leq R_H^{\gamma \gamma} \,.
\end{align} \end{subequations}
The first of these constraints places $m_h$ in the range where an
excess of events had been observed at LEP \cite{leph}. Similarly,
(\ref{h_cons}b) ensures that $m_H$ agrees with the value reported by
the LHC experiments \cite{LHC_Higgs}. In both cases, the range is a
crude estimate of theoretical and experimental uncertainties. Note
that the peak at the LHC is somewhat narrower than at at LEP, since
the latter has been observed chiefly in multi--hadron final states.

The third constraint \cite{D1} ensures that the $Zh$ production cross
section at LEP is roughly ten times smaller than the corresponding
cross section in the SM, for given mass of the Higgs boson; as noted
above, the excess at LEP is compatible with Higgs production only if
the $ZZh$ coupling is suppressed.

The last two conditions ensure that the LHC signals in the $W W^*$,
four lepton and di--photon channels come out roughly correct. They are
described by the quantities
\be \label{R}
R_H^{XX} \equiv \frac { \Gamma(H \rightarrow gg) } { \Gamma(H_{\rm SM}
  \rightarrow gg) } \cdot \frac { \Gamma(H \rightarrow XX) } { \Gamma(H_{\rm
    SM} \rightarrow XX) } \cdot \frac { \Gamma(H_{\rm SM, \, tot}) }
{ \Gamma(H_{\rm tot}) }.
\ee
They describe the strength of the $H$ signal in the $XX$ channel
normalized to the strength of the corresponding signal for the SM
Higgs boson $H_{\rm SM}$. Here I have assumed that Higgs production at
the LHC is dominated by gluon fusion; this is true both in the SM and
in the relevant parameter range of the MSSM. The strength of the four
lepton signal observed at the LHC, which in the Higgs interpretation
of the signal is due to the decay of the Higgs boson into a real and a
virtual $Z$ boson, agrees quite well with the SM prediction; I
therefore allow this signal to be at most a factor of two stronger or
weaker than in the SM. In contrast, the di--photon signal appears
somewhat stronger than in the SM; I therefore only impose a lower
bound on the strength of the signal in this channel. Since the $\gamma
\gamma$ invariant mass peak has a finite width of roughly 1 GeV,
essentially given by the experimental resolution, $R_H^{\gamma
  \gamma}$ includes the contribution from $gg \rightarrow A
\rightarrow \gamma \gamma$ whenever $|m_H - m_A| < 1$ GeV. However, in
the parameter region of interest this contribution is always very
small, due to the small branching ratio of $A \rightarrow \gamma
\gamma$ decays, which in turn is due to the absence of an $A W^+ W^-$
coupling. Note also that in the MSSM, $R_H^{WW} = R_H^{ZZ}$, so that
no independent constraint can be imposed on the strength in the
di--lepton channel.

Finally, null results of additional searches for Higgs bosons have to
be imposed. In particular, for charged Higgs bosons with mass (well)
below $m_t - m_b$, ATLAS searches for $t \rightarrow H^+ b$ decays,
with $H^+ \rightarrow \tau^+ \nu_\tau$, exclude \cite{atlas_top}
both small and large values of $\tan\beta$, leaving an allowed strip
centered at $\tan\beta \simeq \sqrt{m_t(m_t)/m_b(m_t)} \simeq 7$ where
the $H^+tb$ coupling is minimal. At least in the present context, the
CMS search for neutral MSSM Higgs bosons in the di--tau channel
\cite{cms_tau} is even more constraining. Here I have taken both
analyses at face value. Since the ATLAS charged Higgs search is
basically independent of the details of the neutral Higgs spectrum, it
should indeed apply to the present scenario. CMS states its bounds on
MSSM parameter space in the context of the ``maximal mixing''
scenario, which maximizes $m_h$ for given average stop mass. In this
scenario the CP--odd state is typically quite closely degenerate with
either $h$ or $H$, especially for large $\tan\beta$ where this search
is most sensitive. Such a degeneracy obviously increases the yield of
tau pairs of a given invariant mass. In the present context the mass
splittings between all three neutral Higgs bosons are often sizable;
this should lead to somewhat smaller signals in the di--tau channel
than in the ``maximal mixing'' scenario. Incorporating the CMS
constraints in the $(m_A, \tan\beta)$ plane without modification
therefore probably overstates their impact somewhat. However, this is
not easy to quantify without a full simulation including experimental
resolutions.

This concludes the description of the analysis. Let us now turn to the
results.

\section{Results}
\setcounter{footnote}{0}

This Section contains a discussion of the results of the scan of
parameter space, subject to the constraints discussed in the previous
Section. Of course, the first and quite nontrivial result is that
allowed sets of parameter sets can indeed be found, i.e. the
(phenomenological) MSSM can indeed explain at the same time the (mild)
excess of Higgs--like events at LEP and the detection of a Higgs--like
particle by the LHC experiments.

In order to further test this scenario, one has to know what it
implies for the relevant observables. To that end, I will first
describe upper and/or lower bounds on quantities of interest that were
found in the scan, before discussing correlations between pairs of
these quantities.

\subsection{Bounds on Observables}

Let us first look at observables in the Higgs sector. Note first of
all that the upper and lower limits on both the $h$ and $H$ mass can
be saturated, i.e. the scenario doesn't allow to further shrink either
of these mass regions beyond the limits imposed as constraints in
eqs.(\ref{h_cons}a,b).

However, not surprisingly there {\em are} nontrivial bounds on the
masses of the CP--odd and charged Higgs bosons. Some bounds already
follow \cite{D1} from the constraint (\ref{h_cons}c) on the $Zhh$
coupling: if $m_A$ or $m_{H^+}$ becomes very large, $h$ automatically
becomes SM--like; in this ``decoupling scenario'' the upper bound on
the $Zhh$ coupling is therefore badly violated. At the same time the
constraint $m_H > 123$ GeV imposes a non--trivial lower bound on the
mass of the charged Higgs. Altogether I find
\be \label{Hplus}
120 \ {\rm GeV} \leq m_{H^+} \leq 170 \ {\rm GeV}\,.
\ee
The upper bound can be saturated, implying that $t \rightarrow H^+ b$
decays can be closed kinematically. This is in (mild) conflict with a
statement of \cite{kajae}, probably due to the large range of
parameters I explored here. Saturating this upper bound requires very
large $\mu$, a large hierarchy between the $\tilde t_L$ and $\tilde
t_R$ masses, a top mixing parameter $|A_t|$ saturating its upper bound
(with $A_t < 0, \, \mu > 0$), and moderate $\tan\beta \simeq 6$.

The corresponding allowed range for the mass of the CP--odd Higgs
boson $A$ reads
\be \label{A}
96 \ {\rm GeV} \leq m_A \leq 152 \ {\rm GeV}\,.
\ee
The upper bound on $m_A$ is saturated for the same choice of
parameters as the upper bound on $m_{H^+}$. The lower bound on $m_A$
together with the constraint (\ref{h_cons}a) on $m_h$ implies that
limits from searches for $hA$ production at LEP are always satisfied.

The LHC searches for non--SM Higgs bosons discussed at the end of
Sec.~2 considerably restrict the allowed values of $\tan\beta$,
leading to
\be \label{tanb}
5.5 \leq \tan\beta \leq 12.5\,.
\ee
The lower bound is largely determined by the ATLAS search for charged
Higgs bosons, while the upper limit is chiefly due to the CMS search
for neutral Higgs bosons in the di--tau channel. The allowed range of
$\tan\beta$ thus looks quite narrow. However, closing it entirely may
not be easy. As noted above, the $H^+tb$ coupling reaches its minimum
near $\tan\beta = \sqrt{m_t/m_b}$, which falls in the range
(\ref{tanb}). The signal strength in the di--tau channel scales
essentially like $\tan^2\beta$, so reducing the upper bound on
$\tan\beta$ by a factor of about 2.5 requires an increase of the
sensitivity of the search by a factor of six. Recall also that my
interpretation of the CMS bound might be overly strict, i.e. the true
bound might be somewhat weaker.

As noted earlier, the constraint (\ref{h_cons}c) implies that the
$HWW$ and $HZZ$ couplings have close to SM strength. However, this
doesn't imply that the $gg \rightarrow H \rightarrow Z Z^* \rightarrow
4 \ell$ signal also has close to SM strength. On the one hand, loops
of new strongly interacting sparticles, in particular stops, can
change the $H$ production cross section significantly. On the other
hand, the couplings of $H$ to SM fermions, in particular to $b$ quarks
and $\tau$ leptons, can still differ considerably from their SM
values, thereby modifying the $H$ decay branching ratios. As a result,
both the upper and the lower limits on $R_H^{ZZ}$ in (\ref{h_cons}d)
can be saturated, if $\tilde t_1$ is not too heavy. The lower bound on
$R_H^{\gamma \gamma}$ can also be saturated, and the upper bound is
\be \label{RHgg}
R_H^{\gamma \gamma} \leq 2.2\,.
\ee
A significant enhancement of the $H \rightarrow \gamma \gamma$ signal,
which is hinted at by present data, is thus possible in this
scenario. However, this enhancement is mostly due to the increase of
the $H$ production cross section and/or decrease of its total width;
both these effects also increase $R_H^{ZZ}$. In fact, when considering
the ratio of signal strengths\footnote{Such ratios have very recently
  also been discussed in \cite{mullah}, which however assumes that the
  LHC signals are due to the production of $h$, not $H$.} in the
$\gamma \gamma$ and $4 \ell$ (or, more generally, $V V^*$) channels
normalized to their respective SM values, only a moderate deviation
from unity is possible in this scenario:
\be \label{ratio}
0.66 \leq \frac {R_H^{\gamma \gamma}} {R_H^{ZZ}} \leq 1.3 \,.
\ee
I do not find any scenarios where the branching ratio for $H
\rightarrow \gamma \gamma$ decays is affected significantly by $\tilde
\tau$ loops; this is probably due to the vacuum stability constraint
\cite{CCB} $|\mu| \leq 3 (m_{\tilde \tau_L} + m_{\tilde \tau_R})/2$,
which has not been imposed in refs.\cite{stauloop} and
\cite{kajae}. The contribution of charged Higgs loops to this
branching ratio is also always very small, although the charged Higgs
boson is quite light in this scenario, as shown in (\ref{Hplus}).

The di--tau channel is currently poorly constrained by the data. In
fact, the strength of the signal in this channel can deviate quite
significantly from its SM value in the present scenario:
\be \label{RHtau}
0.2 \leq R_H^{\tau\tau} \leq 5.7\, .
\ee
Note that the size of the $\tau$ Yukawa coupling exceeds its SM value
for $\tan\beta > 1$. However, the size of the $H \tau^+ \tau^-$
coupling also depends on the mixing angle $\alpha$ between the neutral
CP--even Higgs bosons, and even vanishes if $\cos\alpha = 0$. This
limit cannot be realized in the present scenario, but a substantial
suppression of the di--tau signal strength is possible. On the other
hand, the maximal enhancement of this signal occurs when stop and
sbottom loops simultaneously enhance the $gg \rightarrow H$ production
cross section and suppress the $H b \bar b$ coupling, while the $H
\tau^+ \tau^-$ coupling is enhanced since $|\cos\alpha| >
|\cos\beta|$. Moreover, if $|m_H - m_A| \leq 4$ GeV, the contribution
$R_A^{\tau\tau}$ has been added to $R_H^{\tau\tau}$, since the $H$ and
$A$ di--tau signals would be difficult to distinguish experimentally
in this case. However, the $A \rightarrow \tau^+ \tau^-$ signal is
quite small in the relevant region of parameter space, which has
$\tan\beta \simeq 6$; here the $\tan\beta$ enhancement of the $A b
\bar b$ coupling cannot yet compensate for the $\cot\beta$ suppression
of the $A t \bar t$ coupling, leading to an $A$ production cross
section from gluon fusion which is significantly smaller than the
corresponding SM value.\footnote{In general there is also a
  significant contribution to the inclusive $A$ production cross
  section from the tree--level process $gg \rightarrow b \bar b A$;
  however, the presence of two additional $b-$jets should allow to
  discriminate this process from SM Higgs production, so this
  contribution should not simply be added to $R_H^{\tau\tau}$ even if
  $m_A = m_H$.} Nevertheless the upper end of the range (\ref{RHtau})
is probably already disfavored by present data, given the absence of a
clear signal in this channel. Note, however, that requiring
$R_H^{\tau\tau} < 3$ does not appreciably alter the allowed ranges of
most other observables.

The di--tau channel also seems to offer the best chance for detecting
the light CP--even scalar $h$ at the LHC in this scenario. The
$\gamma \gamma$ signal is very weak for this state, $R_h^{\gamma
  \gamma} \leq 0.035$; this is due to the reduction of the $h W^+ W^-$
coupling implied by the constraint (\ref{h_cons}c). On the other hand,
the $gg \rightarrow h$ production cross section need not be suppressed
relative to its SM value, so that
\be \label{Rhtau}
0.12 \leq R_h^{\tau\tau} \leq 3.4\,.
\ee
Even the upper end of this range might be difficult to probe at the
LHC, since $h$ is quite close in mass to the $Z$ boson which yields a
much stronger signal in the $\tau^+\tau^-$ channel.  For this reason,
the lower end will probably remain unobservable even for LHC upgrades.

Before concluding this Subsection, let me briefly mention some
constraints on quantities related to the stop sector; these are the
only MSSM parameters not directly related to the Higgs sector for
which some non--trivial constraints can be derived in the present
context. For example, requiring that the heavy CP--even Higgs boson
$H$ explains the LHC signals makes it quite difficult to find
acceptable scenarios with small values of the Higgs(ino) mass
parameter $\mu$: only for $\tan\beta \gsim 9$ do some solutions with
$|\mu| < 400$ GeV survive; for $\tan\beta=10$, some solutions with
$\mu > 0$ saturate the lower bound $|\mu| = 100$ GeV. Similarly,
$|A_t|$ has to exceed 400 GeV, and the sum $|A_t| + |\mu| > 2$ TeV.

Turning to sfermion masses, the $\tilde t_1, \, \tilde b_1$ and
$\tilde \tau_1$ masses can all saturate the lower bounds of 100 GeV;
moreover, no meaningful upper bounds on these masses can be
derived. On the other hand, the $\tilde t_2$ mass must exceed 600 GeV
in this scenario, and the sum of $\tilde t_1$ and $\tilde t_2$ masses
must exceed 900 GeV. For comparison: demanding $m_h > 123$ GeV, as
required if the recent LHC discovery is to be interpreted in terms of
the production and decay of $h$, leads to the lower bound $m_{\tilde
  t_1} + m_{\tilde t_2} > 950$ GeV. This indicates that the amount of
finetuning required in these two MSSM explanations of the LHC signal is
comparable.

\subsection{Correlations between Observables}

\begin{figure}[h!]
\centering
\includegraphics*[width=9.6cm]{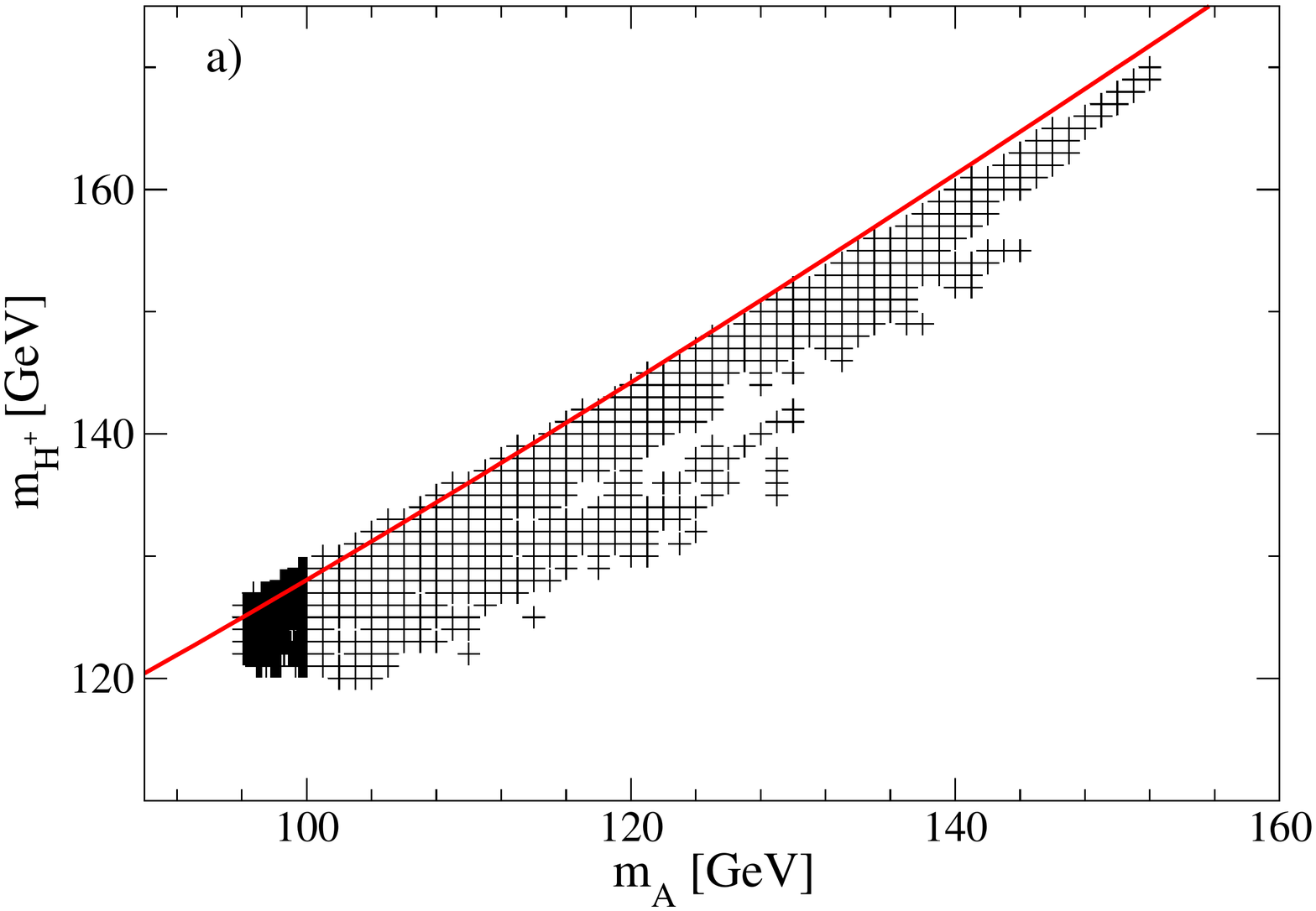} \hspace*{-1.5cm}
\includegraphics*[width=9.6cm]{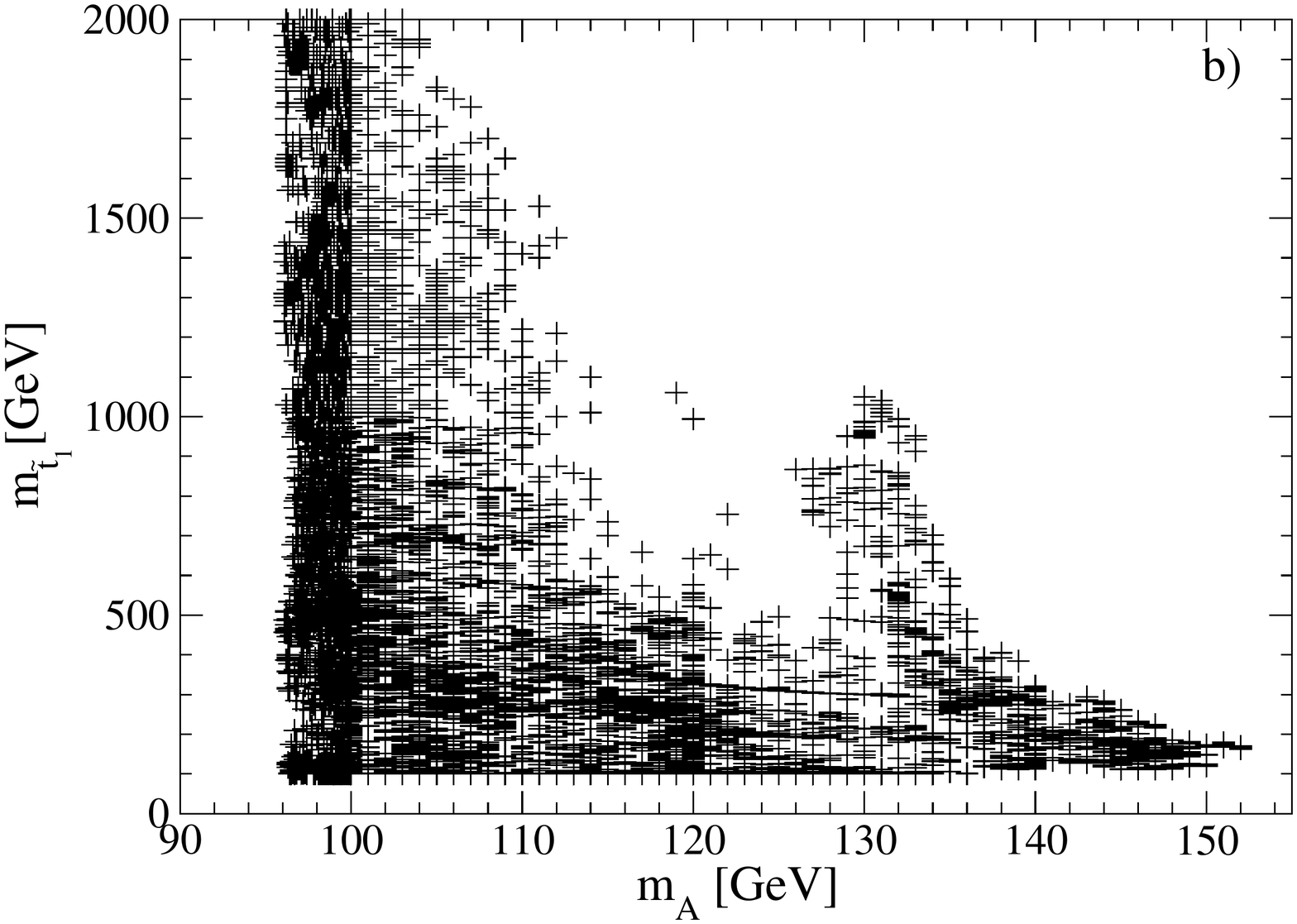}
\caption{%
Allowed region in the $(m_A, m_{H^\pm})$ (a) and the $(m_A,m_{\tilde
  t_1})$ plane (b), after the constraints (\ref{h_cons}) as well as
the  various sparticle and Higgs search limits discussed in the text
have been imposed. The solid (red) line in (a) shows the tree--level
relation between $m_{H^+}$ and $m_A$.}
\label{mm}
\end{figure}

Let us now analyze correlations between observables that result from
the constraints (\ref{h_cons}) as well as the upper limits on MSSM
Higgs searches at the LHC, beginning with correlations between
physical masses. The most obvious such correlations are shown in
Figs.~\ref{mm}a,b, which show the correlation between the mass of the
CP--odd Higgs boson and the mass of the charged Higgs boson and
lighter stop eigenstate, respectively. These, and all following,
scatter plots are based on scans over parameter space containing
several million sets of parameters (not all of which are plotted),
with special emphasis on those regions of parameter space where an
observable reaches an extremum.

As shown in Fig.~\ref{mm}a, the masses of the charged and CP--odd
Higgs bosons are strongly correlated. This is not surprising, given
the tree--level relation $m_{H^+} = \sqrt{m_A^2 + M_W^2}$, which is
indicated by the solid red line. We see that the radiative corrections
to this relation are usually negative, but rather modest in size in
the allowed region of parameter space. This is a consequence of the
upper bounds (\ref{cons}e,f) on the parameters determining stop
mixing, which also largely determine the size of trilinear couplings
between Higgs bosons and stop and sbottom squarks; the upper bound
(\ref{tanb}) on $\tan\beta$ also plays a role in limiting the size of
the corrections to this relation.

The right frame in Fig.~\ref{mm} shows that an {\em upper} bound on
the $\tilde t_1$ mass results in the present scenario if $m_A \gsim
110$ GeV; this results from the upper bound on $m_H$. Notice that for
relatively light $\tilde t_1$, the heavier CP--even Higgs boson can be
significantly lighter than the CP--odd Higgs boson; in contrast, at
the tree level one has $m_H > m_A$. The magnitude of these negative
corrections to $m_H$ is limited by the upper bound on $|A_t|$ and
$|\mu|$ given in (\ref{cons}e).

\begin{figure}[h!]
\centering
\includegraphics*[width=9.6cm]{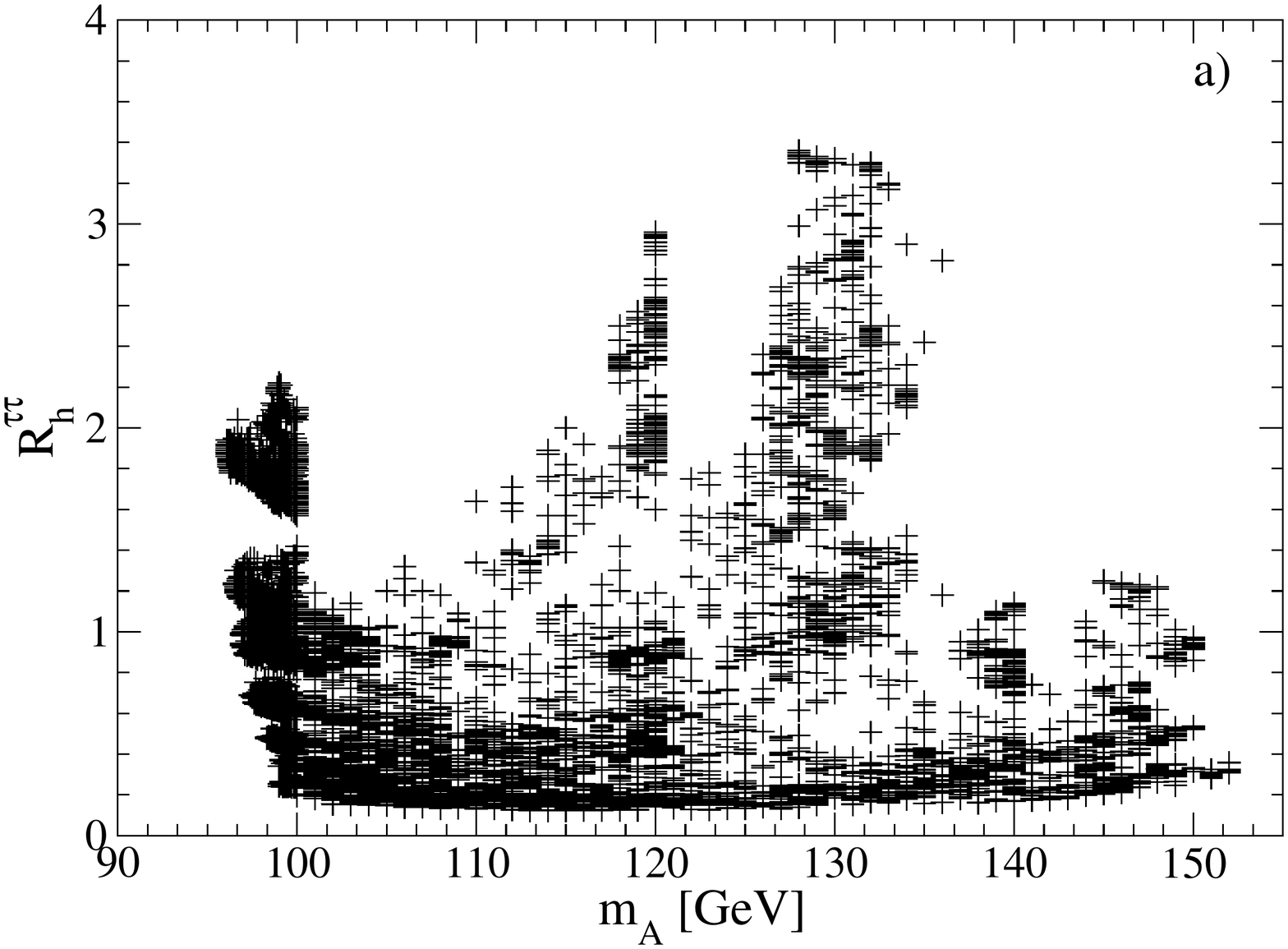} \hspace*{-1.5cm}
\includegraphics*[width=9.6cm]{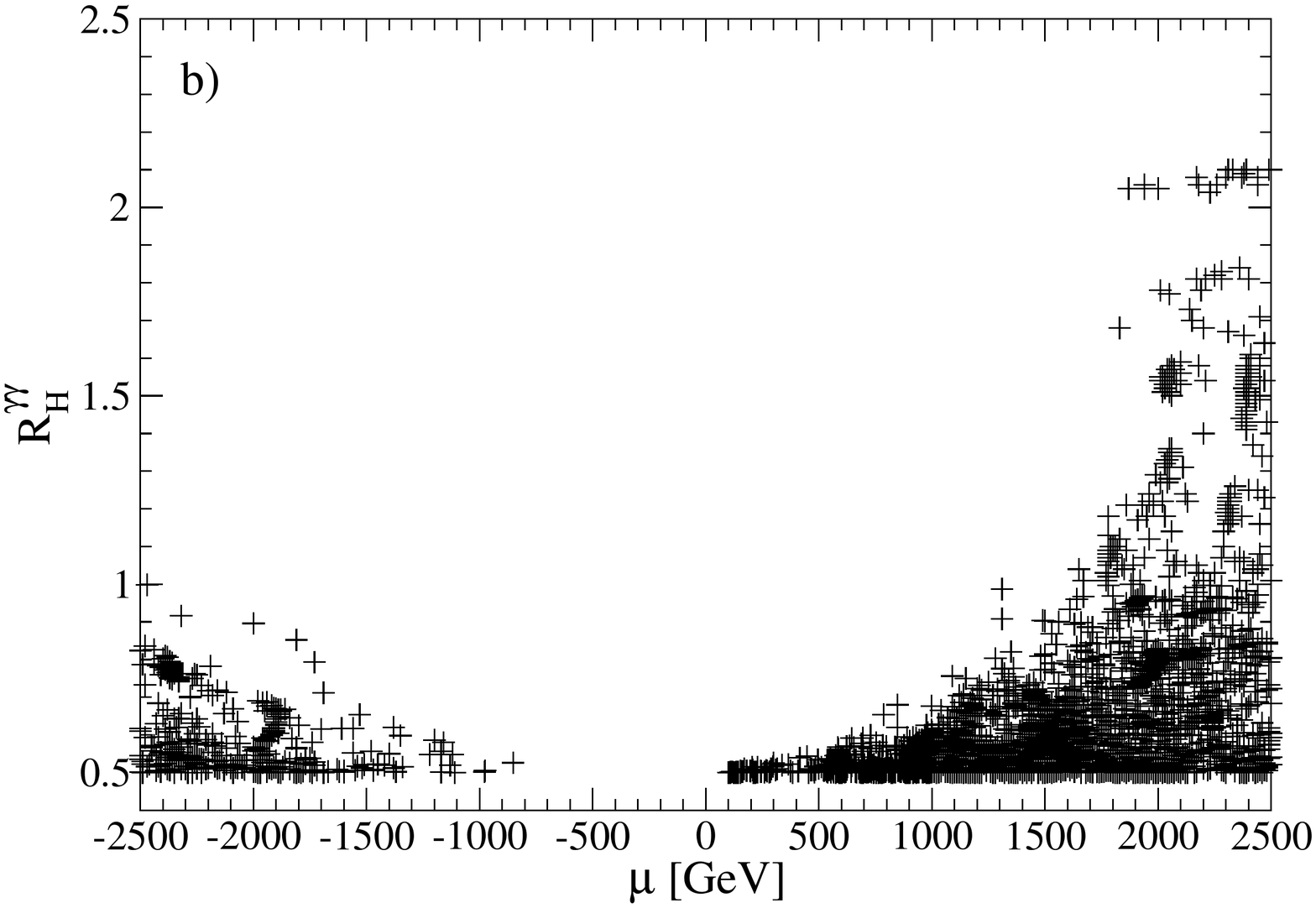} \\
\includegraphics*[width=9.6cm]{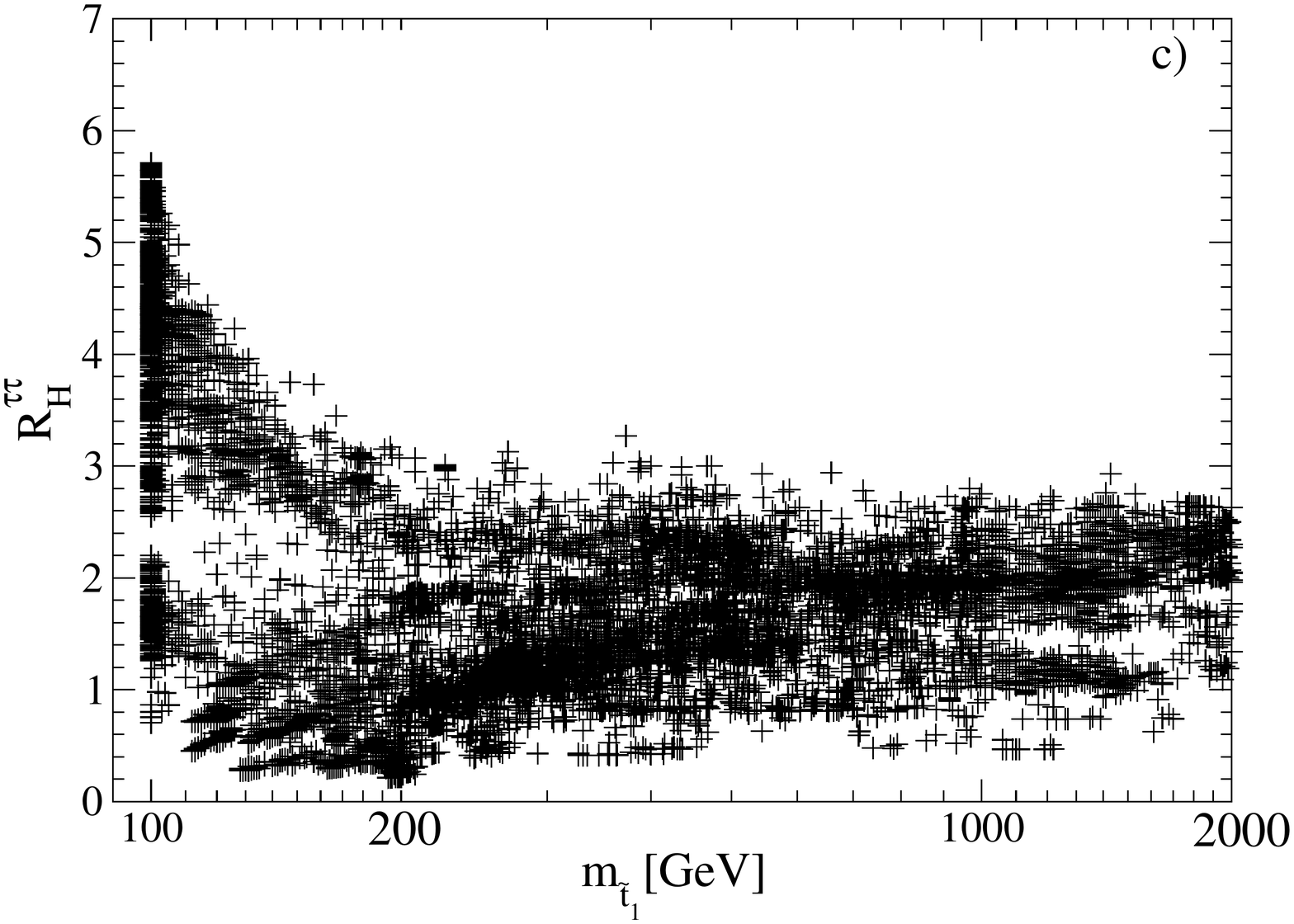} \hspace*{-1.5cm}
\includegraphics*[width=9.6cm]{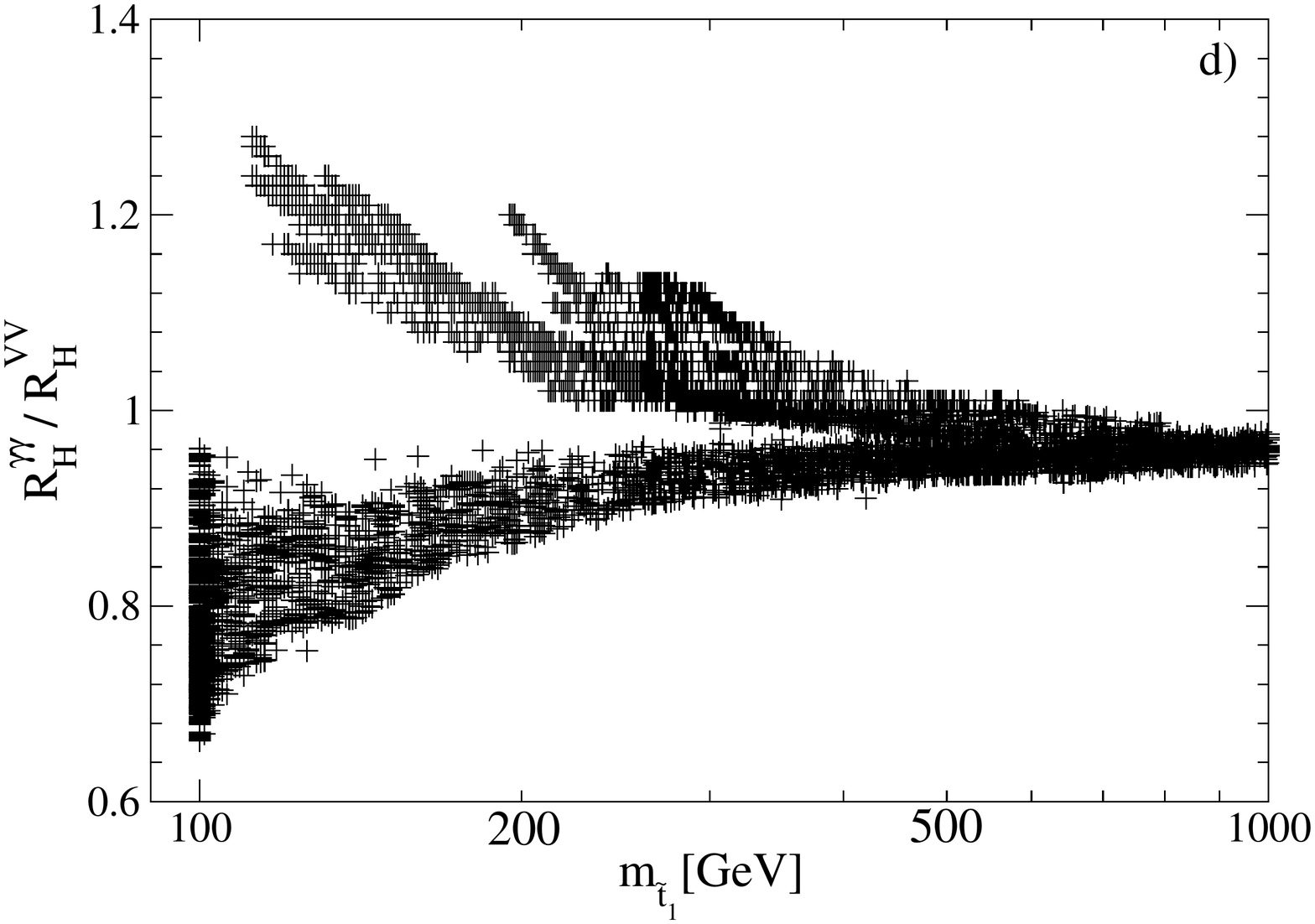}
\caption{%
Allowed region in the $(m_A, R_h^{\tau\tau})$ (a), the
$(\mu,R_H^{\gamma \gamma})$ (b), the $(m_{\tilde t_1},R_H^{\tau\tau})$
(c) and the $(m_{\tilde t_1},R_H^{\gamma\gamma} / R_H^{VV}$) plane (d),
after the constraints (\ref{h_cons}) as well as the  various sparticle
and Higgs search limits discussed in the text have been imposed.}
\label{mr}
\end{figure}

Figure~\ref{mr} shows correlations between a mass and a (ratio of)
signal strength(s). Recall that only the gluon fusion contribution to
various Higgs signals is included here; this should be the dominant
channel in general, but, depending on the cuts, there might be
significant contributions also from $WW$ and $ZZ$ fusion. Associate
production with a $b \bar b$ pair, which can become quite important at
large $\tan\beta$ \cite{abdel}, is not expected to be very important,
given the upper bound (\ref{tanb}).

Frame (a) shows correlations between the mass of the CP--odd Higgs
boson and the $h$ signal strength in the $\tau^+\tau^-$ channel. The
$h \rightarrow \tau^+ \tau^-$ signal strength shows a first peak at
$m_A = 100$ GeV $\simeq m_h$; due to this near--degeneracy, the signal
from $A \rightarrow \tau^+ \tau^-$ has been added, which increases
$R_h^{\tau\tau}$ by up to one unit. This signal reaches its (local)
maximum for the largest allowed value of $\tan\beta$. Here the cross
sections for producing an $h$ or $A$ boson are slightly larger than
the corresponding cross section for producing an SM Higgs boson with
equal mass; the enhancement of the bottom Yukawa coupling
over--compensates the suppression of the couplings of these two
lighter Higgs bosons to top quarks. Both stop squarks need to be
fairly heavy in this region of parameter space, in order to obtain a
sufficiently large value of $m_H$.

Recall that the $A$ and $h$ contributions to this channel are added
only for $|m_A - m_h| \leq 4$ GeV. For intermediate values of $m_A$
the maximal strength of this signal therefore decreases, before
reaching its absolute maximum near $m_A = 130$ GeV. At the absolute
maximum of $R_h^{\tau\tau}$, the lower bound on $m_{\tilde t_1}$ is
saturated, and sbottom loops suppress the $h b \bar b$ coupling. It is
still significantly larger than the corresponding coupling in the SM,
but the enhancement of the $h \tau^+ \tau^-$ coupling is even larger,
leading to an enhanced branching ratio for $h \rightarrow \tau^+
\tau^-$. Moreover, the $h \rightarrow gg$ width, and hence the $h$
production cross section, is dominated by $\tilde t_1$ and $b$ loops,
which have the same sign, while the subleading $t$ loop contribution
has opposite sign. As a result, the gluonic decay width of $h$ exceeds
its SM value by about a factor of two. In combination, this enhances
the $h \rightarrow \tau^+\tau^-$ signal by up to a factor of $3.3$
over its SM value. However, given that $m_h$ is quite close to $M_Z$,
it is not clear whether this enhancement is sufficient to make $h$
detectable at the LHC in this channel. Note also that values of
$R_h^{\tau\tau}$ well below $1$ are possible for nearly all values of
$m_A$. This is chiefly due to the suppression of the $h \rightarrow
gg$ width, which in turn is caused by strong cancellations between the
$t$ and $b$ loop contributions in this region of parameter
space. There is a nontrivial, although phenomenologically probably not
very interesting, lower bound on this quantity, as shown in
(\ref{Rhtau}), since the $b$ loop contribution has a sizable imaginary
part, which cannot be canceled by loops involving much heavier $t$ or
$\tilde t_1$ particles.

Fig.~\ref{mr}b shows that obtaining $|\mu| < 0.5$ TeV is quite
difficult in this scenario. It is possible only if $\tan\beta$ is near
the upper end of the allowed range (\ref{tanb}). Moreover, a large
mass splitting between the $\tilde t_L$ and $\tilde t_R$ masses is
required. The masses of the CP--even Higgs bosons $h$ and $H$ are then
near the lower and upper ends of their allowed ranges,
respectively. In this region of parameter space both the $\tilde t_1$
and the real part of the $b$ loop contributions to $H \rightarrow gg$
as well as $H \rightarrow \gamma \gamma$ have the same sign as the $t$
loop contributions. This enhances the partial width for $H \rightarrow
gg$ by up to a factor of $1.6$, but suppresses the partial width for
$H \rightarrow \gamma \gamma$, which is dominated by $W$ loops, by up
to 20\%. In addition, the partial widths for $H \rightarrow b \bar b$
and $H \rightarrow \tau^+ \tau^-$ are enhanced, further suppressing
the branching ratio for $H \rightarrow \gamma \gamma$. This
overcompensates the increase of the $H$ production cross section.

On the other hand, $R_H^{\gamma \gamma}$ can exceed unity for $|\mu| >
1$ TeV, and reaches its absolute upper bound near $\mu = 2$ TeV. Here
both $\tilde t_1$ and $\tilde b_1$ are quite light, while $m_{\tilde
  t_2} \simeq 1$ TeV, and $\tan\beta \simeq 6$. The light $\tilde t_1$
increases the partial width for $H \rightarrow \gamma \gamma$ by about
5\%, but suppresses the partial width for $H \rightarrow gg$ by about
30\%. This suppression of the total cross section for $H$ production
is over--compensated by the greatly reduced partial widths for $H
\rightarrow b \bar b$ and, to a lesser extent, $H \rightarrow \tau^+
\tau^-$ decays; the light $\tilde b_1$ significantly reduces the $H b
\bar b$ coupling vial SUSY QCD loop corrections in this case. The
signals in the $V V^*$ channels ($V = W^\pm$ or $Z$) are therefore
also enhanced by almost a factor of two.

Note finally that Fig.~\ref{mr}b shows more solutions with $\mu>0$
than with $\mu<0$; moreover, the LEP lower bound on $|\mu|$ can only
be saturated for positive $\mu$. The reason for this asymmetry is that
only positive values of the gluino mass parameter were considered. The
relative sign (more generally, relative phase) between $\mu$ and the
gluino mass parameter has physical meaning, just as the relative signs
(or phases) between $\mu$ and the soft breaking $A-$parameters are
significant. Since only these {\em relative} signs are physical, the
gluino mass parameter can be chosen to be positive without lack of
generality, so long as both signs for $\mu$ and the $A-$parameters are
considered, as is done in the current analysis.

Fig.~\ref{mr}c shows that the $H \rightarrow \tau^+ \tau^-$ signal
strength can be enhanced by more than a factor of three only if
$\tilde t_1$ is very light, $m_{\tilde t_1} \leq 200$ GeV. In this
region of parameter space light $\tilde t_1$ loops increase the $H$
production cross section by about a factor of three over its SM
value. Since $\sin^2(\beta-\alpha)$ saturates its upper bound, the
partial width for $H \rightarrow W W^*$ is reduced by about 15\%,
while the partial widths into the $\tau^+\tau^-$ and $b \bar b$ final
states are increased by factors of $4$ and $2.4$, respectively; the
signals in the $VV$ channels are therefore slightly smaller than in
the SM, in spite of the increased production cross section. Note also
that light $\tilde b_1$ loops again play an important role in
suppressing the $H b \bar b$ coupling.

If all squarks are heavier than a few hundred GeV, the $H \rightarrow
\tau^+ \tau^-$ signal can still be enhanced by up to a factor of
three, essentially by enhancing the $H \tau^+ \tau^-$ couplings since
$|\cos\alpha| > \cos\beta$; since $\tilde b_1$ is heavy, the $H b \bar
b$ coupling will then be enhanced by a similar amount.

Of perhaps greater interest, given current trends in the data, is that
the $H \rightarrow \tau^+ \tau^-$ signal can also be considerably
weaker than in the SM. This signal is weakest for the smallest allowed
value of $\tan\beta$, and requires $\tilde t_1$ and $\tilde b_1$ to be
relatively light; the former reduces the $H$ production rate via gluon
fusion, whereas the latter partly compensates for the reduction of the
$H b \bar b$ coupling that originates from the very small values of
$|\cos \alpha|$ that can be realized in this region of parameter
space. The lower bound on the strength in the $\tau^+ \tau^-$ channel
is then essentially set by the {\em upper} bound (\ref{h_cons}d) on
the strength of the signal in the $VV$ channels, which imposes an
upper bound on the branching ratios for these channels. For larger
squark masses the ratio of the $H \tau^+ \tau^-$ and $H b \bar b$
couplings is essentially fixed, independent of the parameters of the
Higgs sector, leading to a slightly stronger lower bound on the $H
\rightarrow \tau^+ \tau^-$ signal strength. However, even this
increased lower bound is still below conceivable near--future
sensitivities in this channel.

Finally, Fig.~\ref{mr}d shows that the double ratio $R_H^{\gamma
  \gamma} / R_H^{VV}$ can differ by more than $10\%$ from unity only if
$m_{\tilde t_1} < 300$ GeV. Note that the production cross section,
i.e. the partial width for $H \rightarrow g g$, as well as the total
width of $H$ cancel out in this double ratio, which is simply given by
the ratio of the corresponding partial widths, $\Gamma(H \rightarrow
\gamma \gamma) / \Gamma(H \rightarrow W^+ W^-)$, normalized to the
same ratio of partial widths of the SM Higgs boson. Since the $H W^+
W^-$ coupling, which is proportional to $\cos(\alpha-\beta)$, is only
slightly reduced from its SM value, the biggest contribution to
radiative $H \rightarrow \gamma \gamma$ decays always comes from $W$
loops in this scenario. For $m_{\tilde t_1} > 300$ GeV the only other
significant contribution comes from top loops, which always interfere
destructively with the $W$ loops here, just as in the SM. Due to this
destructive interference the reduction of the $H W^+ W^-$ (and $H Z
Z$) coupling implied by the constraint (\ref{h_cons}c) reduces the $H
\rightarrow \gamma \gamma$ partial width slightly more than the $H
\rightarrow V V^* \ (V = W, Z)$ partial widths. However, this
reduction of the double ratio by $\sim 5\%$ will likely remain
unobservable at the LHC.

\begin{figure}[h!]
\centering
\includegraphics*[width=9.6cm]{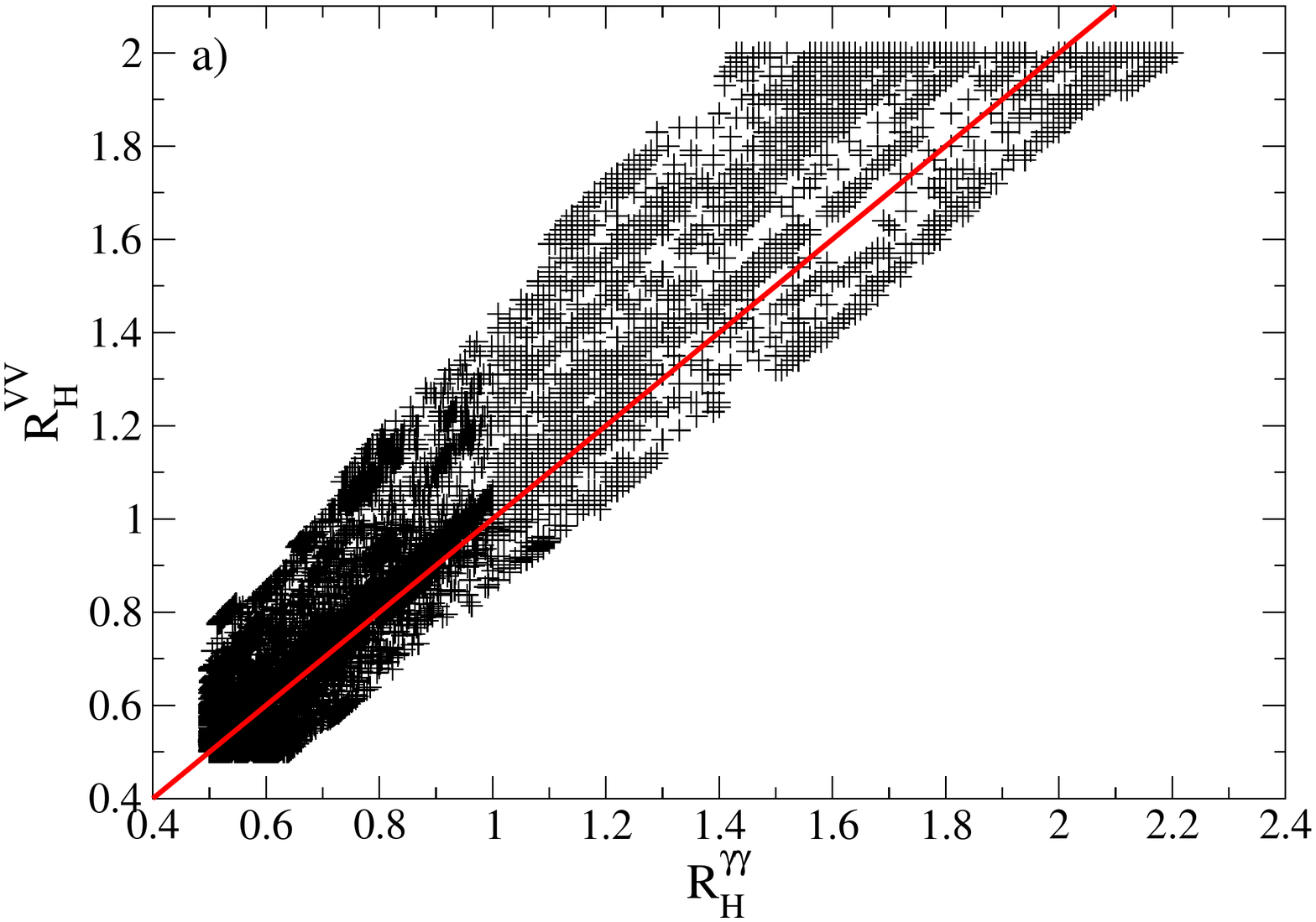} \hspace*{-1.5cm}
\includegraphics*[width=9.6cm]{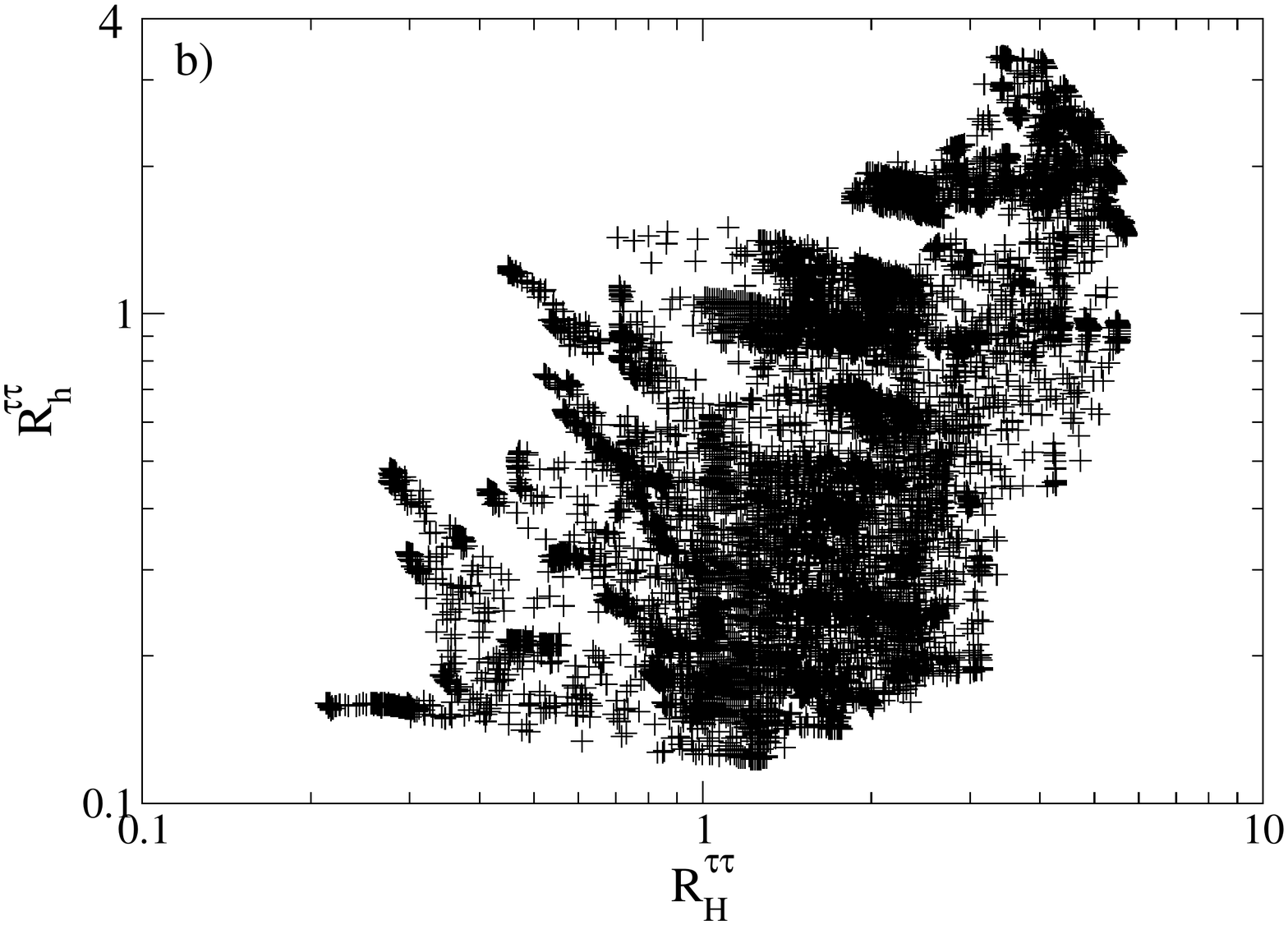} \\
\includegraphics*[width=9.6cm]{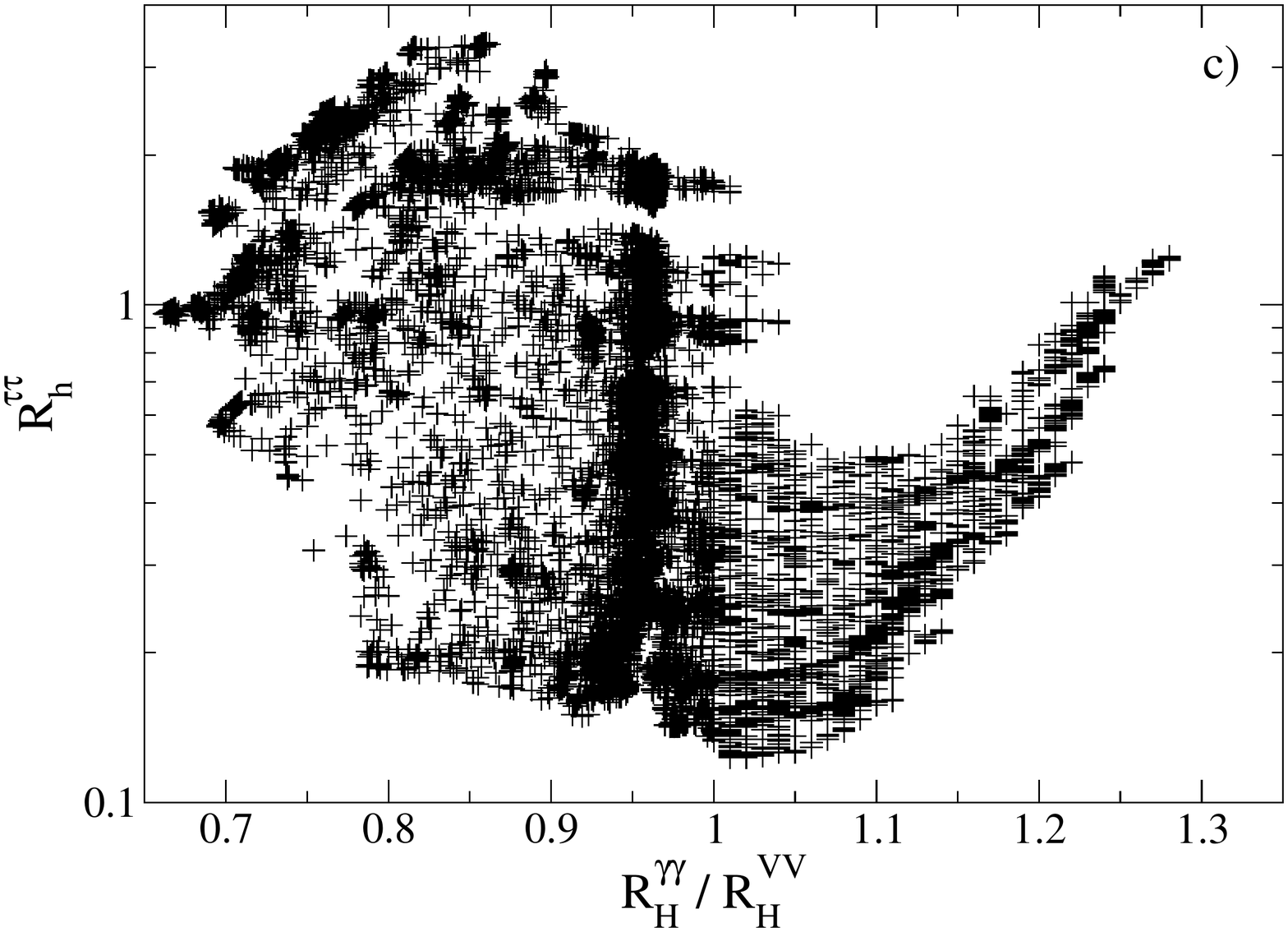} \hspace*{-1.5cm}
\includegraphics*[width=9.6cm]{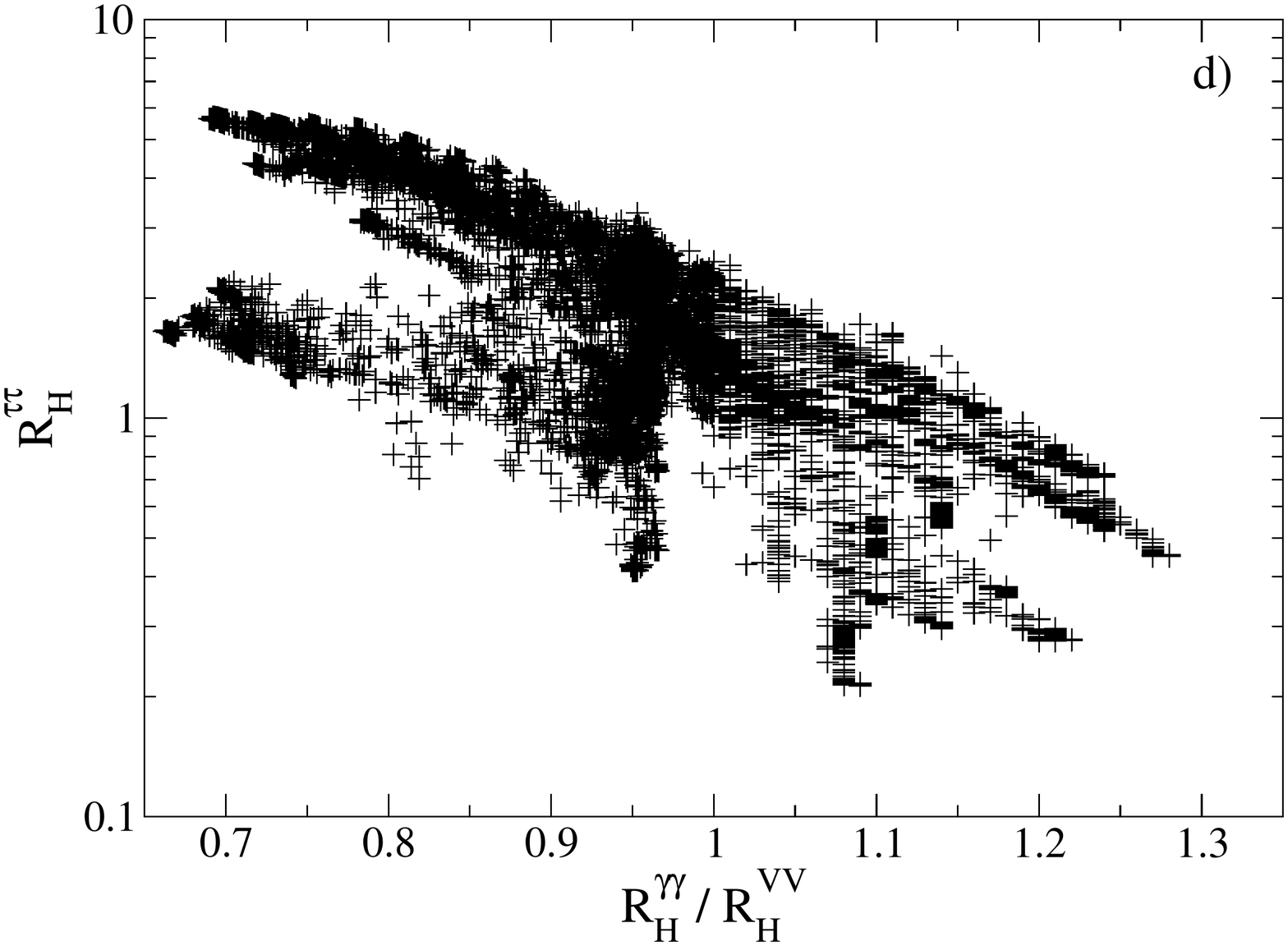}
\caption{%
Allowed region in the $(R_H^{\gamma \gamma}, R_H^{VV})$ (a), the
$(R_H^{\tau\tau},R_h^{\tau \tau})$ (b), the $({\cal R},R_h^{\tau\tau})$
(c) and the $({\cal R},R_H^{\tau\tau})$ plane (d), after the
constraints (\ref{h_cons}) as well as the  various sparticle and Higgs
search limits discussed in the text have been imposed; here ${\cal R}
= R_H^{\gamma \gamma} / R_H^{VV}$. The solid (red) line in frame a)
shows $R_H^{\gamma\gamma} = R_H^{VV}$. } 
\label{rr}
\end{figure}

For smaller $\tilde t_1$ mass the double ratio may differ by up to
$\sim 30$\% from unity. This is due to the effect of $\tilde t_1$ and,
to a lesser extent, $\tilde b_1$ loops on $\Gamma(H \rightarrow \gamma
\gamma)$; recall that equal soft breaking parameters have been used in
the $\tilde t$ and $\tilde b$ sectors here. Depending on the sign of
the $H \tilde t_1 \tilde{t}_1^\dagger$ coupling, this contribution can
interfere constructively or destructively with the dominant $W$ loop
contribution; this explains the bifurcation of the results for
$m_{\tilde t_1} < 400$ GeV. In either case the effect is maximized for
large mass splitting between the $\tilde t$ mass eigenstates and large
$|\mu|$, with maximal suppression (enhancement) of the double ratio
requiring positive (negative) $\mu$. A light $\tilde \tau_1$ can
suppress the double ratio by an additional 3\% or so.

The correlation between the two signal rates defining the double ratio
is explored in the first frame of Fig.~\ref{rr}, which shows
correlations between various (ratios of) signal strengths. The red
line corresponds to $R_H^{\gamma \gamma} = R_H^{VV}$, leading to a
unit value for the double ratio. We see that both signal strengths can
saturate their lower bounds defined in (\ref{h_cons}d,e), and that
$R_H^{VV}$ can also saturate its upper bound. Evidently relaxing the
bounds on $R_H^{VV}$ would also lead to an increased allowed range for
$R_H^{\gamma \gamma}$, beyond the range shown in (\ref{RHgg}). On the
other hand, neither $R_H^{\gamma \gamma}$ nor $R_H^{VV}$ is strongly
correlated with the ratio between these two quantities: the allowed
range of the ratio of signal strengths moves only slightly towards
smaller values as $R_H^{VV}$ increases, such that $R_H^{\gamma \gamma}
\lsim 1.1 R_H^{VV}$ when $R_H^{VV}$ saturates its upper bound of 2;
this is to be compared with the absolute upper bound of $1.3$ on the
ratio, see eq.(\ref{ratio}).

Fig.~\ref{rr}b shows that the $h$ and $H$ signals in the di--tau
channel are positively correlated. The reason is that increasing
$\tan\beta$ increases the basic $\tau$ Yukawa coupling in the
Lagrangian, which therefore also tends to increase the couplings of
both $h$ and $H$ to $\tau$ leptons. However, for small value of
$R_h^{\tau\tau}$ this correlation is not particularly strong: the $H
\rightarrow \tau^+ \tau^-$ signal can then be both significantly
stronger and significantly weaker than in the SM. On the other hand,
the $h \rightarrow \tau^+ \tau^-$ signal strength can only exceed that
of the SM significantly if the $H \rightarrow \tau^+ \tau^-$ signal is
also enhanced. Current data disfavor an enhanced signal in the di--tau
channel for the new boson near 125 GeV; in the present context a
significant upper bound on this signal would make it even more
difficult to detect $h$ at the LHC.

The correlation between the $h \rightarrow \tau^+ \tau^-$ signal
strength and the double ratio $R_H^{\gamma \gamma} / R_H^{VV}$ is
explored in Fig.~\ref{rr}c. We see that the former can only be
enhanced significantly beyond its SM value if the latter is somewhat
below unity. Recall from the discussion of Fig.~\ref{mr}a that
maximizing $R_h^{\tau\tau}$ requires small $m_{\tilde t_1}$. In the
case at hand this enhances the partial width for $H \rightarrow g g$,
but reduces the partial width for $H \rightarrow \gamma \gamma$,
leading to a reduction of the ratio of signal strengths in the $\gamma
\gamma$ and $V V^*$ channels relative to their SM value. Again the
current data favor this double ratio to be enhanced; Fig.~\ref{rr}c
shows that this would reduce the upper bound on the $h \rightarrow
\tau^+ \tau^-$ signal strength in this scenario.

Finally, Fig.~\ref{rr}d shows the correlation between the double ratio
of $H \rightarrow \gamma \gamma$ and $H \rightarrow V V^*$ signal
strengths and the $H \rightarrow \tau^+ \tau^-$ signal. These
quantities are clearly anti--correlated in the present scheme. The $H
\rightarrow \tau^+ \tau^-$ signal is maximized in a similar region of
parameter space as the $h \rightarrow \tau^+ \tau^-$ signal; we just
saw that this leads to a suppression of the double ratio.

Conversely, the double ratio reaches its maximum when both the $H
\rightarrow \gamma \gamma$ and $H \rightarrow V V^*$ signals are
suppressed by destructive interference of top and stop loop
contributions to $H \rightarrow gg$; stop loop contributions then
maximally enhance $\Gamma(H \rightarrow \gamma \gamma)$. The
suppression of the $H$ production cross section also reduces the
strength of the signal in the di--tau channel. One can also find
configurations with slightly less enhanced double ratio where both the
$H \rightarrow \gamma \gamma$ and the $H \rightarrow V V^*$ signals
are enhanced over their SM values. This can be achieved if $|\cos
\alpha| < \cos\beta$, which suppresses the $H \rightarrow b \bar b$
and $H \rightarrow \tau^+ \tau^-$ partial widths, and thus the total
decay width of $H$. This mechanism also leads to a suppression of the
$H \rightarrow \tau^+ \tau^-$ signal. Finally, the branch with
$R_H^{\gamma \gamma} / R_H^{VV} \simeq 1.2, \, R_H^{\tau\tau} \simeq
0.3$ requires very large and negative $\mu$, large and positive $A_t$,
and (as usual) large splitting between $m_{\tilde t_L}$ and $m_{\tilde
  t_R}$. The light $\tilde t_1$ loops then again suppress $H
\rightarrow gg$ decays and enhance $H \rightarrow \gamma \gamma$
decays, whereas $|\cos \alpha| < \cos\beta$ reduces the total width of
$H$; the former effect is dominant, i.e. the $H \rightarrow \gamma
\gamma$ and $H \rightarrow V V^*$ signals are both suppressed relative
to their SM values. In this case the di--tau signal is further
suppressed because the light $\tilde b_1$ enhances, rather than
suppresses, the ratio of $H b \bar b$ and $H \tau^+ \tau^-$ coupling.

\section{Summary and Conclusions}

In this paper I have shown that one can explain both the recent
discovery of a ``Higgs--like particle'' by the LHC experiments, and
the $2.3 \sigma$ excess of Higgs--like events found by the LEP
collaborations some ten years ago, in the ``phenomenological'' MSSM,
where all (relevant) weak--scale soft breaking parameters are treated
as independent free parameters. In this interpretation the masses of
the two CP--even Higgs bosons $h$ and $H$ are essentially fixed by the
data, to $\sim 98$ and $\sim 125$ GeV, respectively. Radiative
corrections to the Higgs sector are crucial for the viability of this
scheme. As a result, the masses of the remaining Higgs bosons, the
CP--odd state $A$ and the charged state $H^\pm$, can still vary
considerably. Nevertheless stringent upper bounds on the masses of
these states can be derived, which would be straightforward to test at
an $e^+ e^-$ collider operating at $\sqrt{s} \geq 350$ GeV. Much of
the allowed parameter space can probably also be probed by searches
for these states at the LHC; in particular, $t \rightarrow H^+ b$
decays are open over almost the entire parameter space. However, this
is not sufficient to guarantee that such decays, or other $A$ or
$H^\pm$ production processes, can actually be detected at the LHC.

The upper bound on $m_{H^\pm}$ implies that loops involving the
charged Higgs boson and the $t$ quark will give significant positive
contributions to the partial width for radiative $b \rightarrow s
\gamma$ decays \cite{bg}; if these were the only new contributions the
predicted partial width would exceed the measured value \cite{pdg},
which is quite close to the SM prediction. However, it is well known
that even within the MSSM with minimal flavor violation,
chargino--stop loops can cancel the charged Higgs loops
\cite{bsg_susy}, so a portion of the parameter space (with rather
light $\tilde t_1$) is most likely allowed even in this constrained
scenario. Moreover, as argued in the Introduction, the general MSSM
contains many additional parameters that can be tuned to satisfy
flavor constraints. In particular, a small amount of $\tilde b -
\tilde s$ mixing would lead to large gluino loop contributions to $b
\rightarrow s \gamma$ \cite{or} of either sign.

The light state $h$ is also very difficult to detect at the LHC. It
has greatly reduced couplings to $Z$ and $W$ bosons, and hence also a
greatly reduced branching ratio into $\gamma \gamma$ final
states. Part of the allowed parameter space could perhaps be probed
through $h \rightarrow \tau^+ \tau^-$ decays, but the small value of
$m_h$ implies that $Z \rightarrow \tau^+ \tau^-$ decays will be a
formidable background.

Although the couplings of $H$ to $W$ and $Z$ bosons are quite SM--like
in this scenario, both the production cross section and the decay
branching ratios of $H$ can still differ significantly from those of
the SM Higgs. On the one hand, the couplings of $H$ to third
generation fermions can be quite different from those of the SM Higgs;
here light $\tilde b$ loop contributions to the $H b \bar b$ coupling
can play a significant role. A good measurement of, or upper bound on,
the strength of the di--tau signal would therefore narrow down the
allowed parameter space of this scenario. A reliable observation of
the $H \rightarrow b \bar b$ signal and/or $t \bar t H$ production
would be similarly useful, but are experimentally (even) more
challenging. Moreover, light $\tilde t$ loops can modify the partial
widths for $H \rightarrow gg$ and, to a somewhat lesser extent, for $H
\rightarrow \gamma \gamma$ decays significantly. In particular, in
this scenario one can simultaneously reduce the di--tau signal and
enhance the di--photon signal, in agreement with the (statistically
not very compelling) trend of current data.

However, the di--photon signal can be enhanced relative to the $V V^*
\ (V = W^\pm, Z)$ signals only if $\tilde t_1$ is rather light.
On--going and future searches for light stop and sbottom squarks
therefore have the potential to further constrain the parameter space
of this model. Unfortunately the interpretation of such searches also
depends on the chargino and neutralino sectors of the MSSM, which have
not been specified here, since they hardly affect the Higgs
sector. Note, however, that the LHC experiments should eventually be
able to probe $\tilde t_1$ masses well above 200 GeV even in the
experimentally most difficult case where $\tilde t_1$ is nearly
degenerate with a stable neutralino \cite{stop}. If all squarks are
heavy, $H$ can be so SM--like that it probably cannot be distinguished
from the Higgs boson of the SM. In particular, the squared couplings
to $W$ and $Z$ are just $5$ to $15\%$ smaller than in the SM; this
follows from the normalization of the excess observed at LEP.

It should be admitted that this scenario is theoretically not
especially appealing. In particular, the LEP excess cannot be
explained in a constrained version of the MSSM
\cite{no_cmssm}. Moreover, the lower bound on the sum of the stop
masses, and hence the required finetuning associated with radiative
corrections from the top and stop sector, is similar to that in the
more common MSSM interpretation of the LHC discovery in terms of
production and decay of the light CP--even state $h$. It is
nevertheless amusing to note that the MSSM can {\em simultaneously}
explain two sets of observations of excesses of Higgs--like events.

\subsection*{Acknowledgments}
This work was supported in part by the BMBF--Theorieverbund, and in
part by the DFG Transregio TR33 ``The Dark Universe''.

\end{document}